\documentclass[conference]{IEEEtran}

\usepackage[colorlinks=true, linkcolor=black, citecolor=black, urlcolor=black]{hyperref}
\usepackage[subtle,tracking=normal]{savetrees}
\usepackage{cite}
\usepackage{amsmath,amssymb,amsfonts}
\usepackage{algorithmic}
\usepackage{graphicx}
\usepackage{textcomp}
\usepackage{xcolor}
\usepackage{subcaption}
\usepackage{graphicx}
\usepackage{placeins}
\usepackage{caption}

\def\BibTeX{{\rm B\kern-.05em{\sc i\kern-.025em b}\kern-.08em
T\kern-.1667em\lower.7ex\hbox{E}\kern-.125emX}}

\begin{document}

\title{GNN-based Online Beamforming Design for \\ HAPS-Assisted NTN}
\author{\IEEEauthorblockN{ Lavanya S S Anjapuli\IEEEauthorrefmark{1}, Animesh Yadav\IEEEauthorrefmark{1}, and Halim Yanikomeroglu\IEEEauthorrefmark{2}}
\IEEEauthorblockA{\IEEEauthorrefmark{1}School of EECS, Ohio University, Athens, OH, USA\\ \IEEEauthorrefmark{2}NTN Lab., Department of SCE, Carleton University,  Ottawa, ON, Canada\\
Email: \{ls775924, yadava\}@ohio.edu\IEEEauthorrefmark{1}, halim@sce.carleton.ca\IEEEauthorrefmark{2}}}
\maketitle

\begin{abstract}
In terrestrial networks, especially in urban areas, cell-edge users often face significant capacity limitations due to high path loss, shadowing, and inter-cell interference (ICI). This paper proposes integrating a high-altitude platform station (HAPS) into terrestrial networks, where terrestrial base stations (BS) can alleviate these issues by relaying data intended for cell-edge users via HAPS, thereby leveraging line-of-sight (LoS) links. We formulate an energy-efficiency (EE) maximization problem to jointly design beamforming vectors at the BS and HAPS with the goal of improving cell-edge user performance. Since the resulting problem is non-convex, we develop an online optimization framework based on a graph neural networks (GNN), which effectively captures the network topology. Numerical results show that the proposed HAPS-assisted architecture improves network performance, particularly by increasing the 5th-percentile EE, thereby enhancing service for cell-edge users.
\end{abstract} 

\begin{IEEEkeywords}
Energy efficiency, Beamforming design, graph neural networks (GNN), high altitude platform stations (HAPS), online optimization, non-terrestrial networks (NTN).
\end{IEEEkeywords}

\section{Introduction}

Multi-cell network architectures form the backbone of cellular wireless systems and are continued to remain fundamental in beyond fifth-generation (5G) and sixth-generation (6G) networks to support dense connectivity, high data rates, and wide-area coverage through massive multiple input multiple output (MIMO), advanced beamforming, and dynamic interference management \cite{Andrews-Buzzi-JSAC-2014}. However, inter-cell interference, shadowing, and non-line-of-sight (NLoS) propagation significantly degrade the performance of cell-edge users, limiting spectral and energy-efficiency (EE) \cite{Gesbert-Hanly-JSAC-2010}. High-altitude platform station (HAPS) have therefore emerged as a promising solution to these challenges by offering large-scale coverage with a high probability of line-of-sight (LoS) air-to-ground links, thereby effectively mitigating shadowing effects \cite{Kanani-Parisa-Omidi-Javed-JSAC-2025}.



Furthermore, due to the presence of LoS air-to-ground links and the availability of wide bandwidth, millimeter-wave (mmWave) communication is well suited for HAPS-assisted networks. The LoS links help mmWave signals compensating severe path loss but more can be achieved by employing large antenna elements to achieve beamforming gain \cite{Rappaport-IEEEAccess-2013}. Implementing fully digital beamforming with large antenna arrays is often impractical, as it requires one radio-frequency (RF) chain per antenna element, resulting in high hardware cost, power consumption, and baseband processing complexity \cite{Molisch-Ratnam-HB-2017}. Hybrid beamforming offers a practical alternative by combining analog RF beam steering with low-dimensional digital precoding, thereby providing a favorable trade-off between performance and hardware efficiency. Nevertheless, hybrid architectures introduce additional challenges in beamformer optimization under constant-modulus constraints and dynamic resource allocation, motivating the development of advanced design methodologies \cite{Sun-INFOCOM-2018,Ayach-TWC-2014}.

Iterative optimization-based algorithms for hybrid beamforming design include alternating optimization methods \cite{Yu-Shen-Zhang-Letaief-TSP-2016, Qiao-Zhang-Zhou-Yang-IEEEAccess-2020}, adaptive cross-entropy (ACE) approaches \cite{Hassan-Shabih-Alamari-Sultan-Usama-Electronics-2023}, and stochastic beamforming techniques \cite{Feng-Wenzhi-Fu-SAM-2020}. Recently, learning-based methods have gained traction for their ability to reduce the computational burden of iterative optimization \cite{Sung-Lakju-cho-IEEEAccess-2020, Ayad-Mohammed-Fethi-IJECE-2025, zhang-Teng-Dong-Anming-Electronics-2022, Sun-Qiang-Zhao-Huan-Wang-Entropy-2022}.  For instance, the deep neural network (DNN)-based hybrid beamforming framework in \cite{Sung-Lakju-cho-IEEEAccess-2020} reduces training complexity in multi-user mmWave MIMO systems. \cite{Ayad-Mohammed-Fethi-IJECE-2025} develops a convolutional neural networks (CNN) architecture through supervised training to jointly learn analog and digital beamforming weights for improved energy efficiency in mmWave massive MIMO. Similarly, \cite{zhang-Teng-Dong-Anming-Electronics-2022} presents a CNN model that optimizes hybrid beamforming weights in a multiple-input single-output (MISO) system using an signal-to-noise ratio (SNR) as a loss function. Lastly, \cite{Sun-Qiang-Zhao-Huan-Wang-Entropy-2022} proposes an end-to-end CNN framework that integrates channel state information (CSI) feedback and hybrid beamforming weight design for mmWave MIMO systems, aimed at maximizing spectral efficiency.

While the aforementioned learning-based approaches significantly reduce computational complexity, they typically operate under fixed input–output dimensions and lack explicit modeling of network topology and interference relationships. As a result, their ability to generalize is limited when the number of users or base stations changes. Moreover, practical wireless networks experience time-varying channels due to dynamic propagation environments, which can lead to mismatches between training data and real-time channel distributions. Online optimization frameworks mitigate this issue by continuously updating model parameters using real-time channel observations \cite{chen-guanghui-wang-TWC-2025}. Motivated by these insights, we propose an online graph neural networks (GNN)-based hybrid beamforming framework for energy-efficient HAPS-assisted multi-cell  networks. GNN address these limitations by modeling wireless systems as graphs and explicitly capturing interference relationships, enabling scalable and topology-aware optimization \cite{Shen-Yiefi-Zhang-Jun-Song-Letaief-Khaled-TWC-2022}. The contributions of this
paper are summarized as follow:
\begin{itemize}
\item A GNN-based online optimization technique is developed to jointly optimize hybrid beamforming weights at both terrestrial BS and HAPS by maximizing the EE of the network.
\item Comprehensive performance evaluations is provided by comparing the proposed approach with DNN, CNN and ACE-based techniques demonstrating the superiority of the proposed technique.
\end{itemize}

The remainder of this paper is organized as follows: Section II describes the system model, while Section III formulates the EE optimization problem. The solution to this problem is presented in Section IV. Section V outlines the simulation setup and discusses the results, and Section VI concludes the paper.
\begin{figure}[t]
    \noindent
    \begin{minipage}{1\linewidth}
        \includegraphics[width=\linewidth]{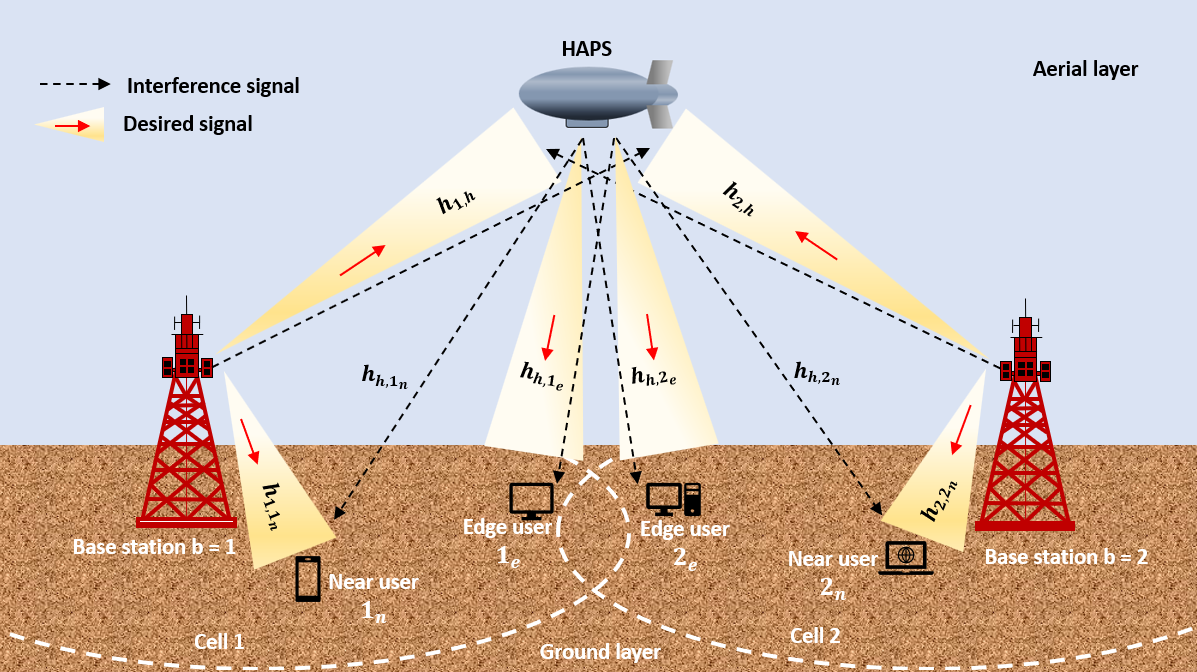}
    \end{minipage}
    \caption{System architecture of the network considered.}
    \label{fig:sys_arch}
\end{figure}
\section{System Model} \label{sec:system_model} 

In this paper, we consider a multi-cell network in which each cell is served by a BS equipped with a uniform planar array (UPA) consisting of $M_b = M_{bh} M_{bv}$ antenna elements where $M_{bh}$ and $M_{bv}$ are the number of elements along the horizontal $x$-axis and vertical $z$-axis, respectively. A full-duplex (FD) HAPS, capable of simultaneous transmission and reception over the same frequency band, is deployed to support cell-edge UE that experience severe inter-cell interference, while near UE are directly served by their associated BS. Specifically, the FD-HAPS simultaneously receives signals from multiple BS and transmits data to edge UE over the same time--frequency resources. \textcolor{black}{The HAPS is also equipped with an UPA} comprising $M_h = M_{hx} M_{hy}$ antenna elements arranged along the horizontal $x$ and $y$ axes. To enhance spectral efficiency, power-domain non-orthogonal multiple access (NOMA) is employed at the transmitting nodes to multiplex near and edge UE signals over the same time–frequency resources, while \textcolor{black}{successive interference cancellation (SIC)} is applied at the receivers to decode the superimposed signals. The entire architecture is illustrated in Fig. \ref{fig:sys_arch}.

All terrestrial links in the considered multi-cell network operate in the millimeter-wave (mmWave) frequency band, typically in the range of 24--100~GHz, to exploit the large available bandwidth for high data rate transmission \cite{Andrews-Buzzi-JSAC-2014,Rappaport-IEEEAccess-2013,Giordani-AdhocNetworking-Madhoc-2016}. In this work, we consider 28 GHz as the operating frequency for all transmissions.

\subsection{Channel Model} \label{sec:channel_model}
Because of the use of mmWave band frequencies, which undergo limited scattering, all the terrestrial links, such as BS-to-UE, are modeled using a sparse geometric multipath framework \cite{Ragnavan-IEEE-TIT-2008}. Considering $S$ dominant scatterers, the channel from BS $b$ to UE $u$ can be represented as
\textcolor{black}
{\begin{eqnarray}
\mathbf{h}_{b, b_u} = \sqrt{\frac{1}{S}}\sum_{k=1}^Sg^k_{b,b_u}\mathbf{r}_t(\theta^{D,k}_{b,b_u}, \varphi^{D,k}_{b,b_u}) \sqrt{G_b\big(\theta^{D,k}_{b,b_u}, \varphi^{D,k}_{b,b_u}\big)}, 
\end{eqnarray}}
where $g^k_{b,b_u}$ represents the complex channel gain of the path $k$, which also includes path loss, \text{PL} = $\left( \frac{4\pi d f_c}{c} \right)^2$ with $f_c$ is the carrier frequency, $c = 3 \times 10^{8}\ \text{m/s}$ is the speed of light in free space and $d$ is the distance from BS to UE \cite{George-Shu-Theodore-Yunchou-Hangsong-MWWC-2016}. The angles $\theta^{D,k}_{b,u}$, $\theta \in [-\pi/2, \pi/2]$ and $\varphi^{D,k}_{b,u}$, $\varphi \in [0, 2\pi]$ are the elevation and azimuth angles of departure, respectively. The term {$G_b\big(\theta^{D,k}_{b,b_u}, \varphi^{D,k}_{b,b_u}\big)$ represents the antenna gain of BS $b$ and $\mathbf{r}_t(\theta^{D,k}_{b,u}, \varphi^{D,k}_{b,u})$, for the path $k$, represents the transmit array response vector at an elevation and azimuth angle of $\theta$ and $\varphi$, which can be defined as
\begin{equation}
    \mathbf{r}_t(\theta, \varphi) = \mathbf{r}_x(\theta, \phi) \otimes \mathbf{r}_z(\theta),
\end{equation}
where $\otimes$ is Kronecker product and $\mathbf{r}_x$ and $\mathbf{r}_z$ are the steering vectors in x-axis and z-axis respectively \cite{Liu-Xuanyu-Shijian-Cheng-LLM4CP-2024}. Considering $d_A$ as the distance between antenna elements and $\lambda$ as the wavelength of the carrier signal, the steering vectors can be defined as
\begin{align}
    \mathbf{r}_x(\theta,\phi) &= \left[1, e^{j\frac{2\pi d_A}{\lambda}\sin\theta\cos\phi},\ldots, e^{j\frac{2\pi d_A}{\lambda}(M_{bh}-1)\sin\theta\cos\phi}\right]^T, \\
    \mathbf{r}_z(\theta) &= \left[1, e^{j\frac{2\pi d_A}{\lambda}\cos\theta},\ldots, e^{j\frac{2\pi d_A}{\lambda}(M_{bv}-1)\cos\theta}\right]^T.
\end{align}


Considering the communication link between the BS and the HAPS, the channel generally consists of a dominant LoS component along with several NLoS components caused by ground-level scatterers such as buildings and terrain. Since the HAPS is located at a high altitude above the ground level, the number of significant multipath components is limited, and thus the BS–HAPS channel can be accurately modeled using a sparse geometric multipath framework with $S$ dominant scatterers.
\begin{flalign}
\mathbf{h}_{b,h} = &\sqrt{\frac{1}{S}}\sum_{k=0}^{S} g^k_{b,h}\,
\mathbf{r}_r(\theta^{A,k}_{b,\text{h}}, \varphi^{A,k}_{b,\text{h}})
\mathbf{r}^{*}_t(\theta^{D,k}_{b,\text{h}}, \varphi^{D,k}_{b,\text{h}}) \notag\\
&\times \sqrt{G_h\big(\theta^{A,k}_{b,\text{h}}, \varphi^{A,k}_{b,\text{h}}\big)\,
G_b\big(\theta^{D,k}_{b,\text{h}}, \varphi^{D,k}_{b,\text{h}}\big)}.
\label{eq:ch_HIBS_2_UE}
\end{flalign}
The first term with $k=0$ in (\ref{eq:ch_HIBS_2_UE}) represents the LoS component and the second term represent the NLoS component. $g^0_{b,\text{h}}$ and $g^k_{b,\text{h}}$ represent the complex channel gain of the LoS and NLoS path $k$, respectively. The terms $\mathbf{r}_r(\theta^{A,k}_{b,\text{h}}, \varphi^{A,k}_{b,\text{h}})$ and $\mathbf{r}_t(\theta^{D,k}_{b,\text{h}}, \varphi^{D,k}_{b,\text{h}})$ represent the receive  and transmit array response vector of NLoS path $k$, respectively. These two terms with $k=0$ represent response vectors of the LoS path respectively. $G_h\big(\theta^{A,k}_{b,\text{h}}, \varphi^{A,k}_{b,\text{h}}\big)$ and $G_b\big(\theta^{D,k}_{b,\text{h}}, \varphi^{D,k}_{b,\text{h}}\big)$ are the antenna gain of HAPS and BS $b$ respectively.

\textcolor{black}{Similar to (\ref{eq:ch_HIBS_2_UE})}, we can define the channel from HAPS to UE  $u$ of the BS $b$ as
\textcolor{black}{\begin{flalign}
\mathbf{h}_{h,b_u} =
\sum_{k=0}^{S} g^k_{h,b_u}\,
\mathbf{r}_t(\theta^{D,k}_{h,{b_u}}, \varphi^{D,k}_{h,{b_u}})
\sqrt{G_h\big(\theta^{D,k}_{h,{b_u}}, \varphi^{D,k}_{h,{b_u}}\big)}
\label{eq:ch_BS_2_HIBS},
\end{flalign}}
where $g^0_{h,{b_u}}$ and $g^k_{h,{b_u}}$ denote the complex channel gains of the LoS and NLoS paths, respectively, and 
$\mathbf{r}_t(\theta^{D,k}_{h,{b_u}}, \varphi^{D,k}_{h,{b_u}})$ represents the transmit array response vector of path $k$ (with $k=0$ corresponding to the LoS path). 
$G_h\big(\theta^{D,k}_{h,{b_u}}, \varphi^{D,k}_{h,{b_u}}\big)$ denotes the antenna gain of the HAPS.

\subsection{Signal Model} \label{sec:signal_model}

In the system considered, each BS serves two UE by transmitting two independent data streams over the same time–frequency resources with different power levels. Hybrid beamforming is adopted at the BS since fully digital beamforming requires one RF chain per antenna, leading to high power consumption for large arrays, while fully analog beamforming relies on a single RF chain and therefore lacks the flexibility required for multi-user transmission and effective inter-user interference mitigation. Hybrid beamforming achieves a practical trade-off between analog and digital beamforming flexibility and interference mitigation capability, and hardware complexity and power consumption, by combining digital and analog processing. Specifically, the data streams are processed using digital beamforming weights
$\mathbf{W}^{\text{D}}_b = [\mathbf{w}^{\text{D}}_{b,n}, \mathbf{w}^{\text{D}}_{b,e}] \in \mathbb{C}^{2 \times 2}$,
followed by analog beamforming weights
$\mathbf{W}^{\text{A}}_b = [\mathbf{w}^{\text{A}}_{b,n}, \mathbf{w}^{\text{A}}_{b,e}] \in \mathbb{C}^{M_b \times 2}$,
where $\mathbf{w}^{\text{D}}_{b,u} \in \mathbb{C}^{2 \times 1}$ denotes the digital beamforming vector for UE $u \in \{n,e\}$ at BS $b$, and $\mathbf{w}^{\text{A}}_{b,u} \in \mathbb{C}^{M_b \times 1}$ denotes the corresponding analog beamforming vector.

At time instant $t$, the signal received at the HAPS is a superposition of four data streams, two transmitted from each BS can be represented as
\begin{equation}
\mathbf{x}_{\text{b,h}}[t] = \sum_{b=1}^2 \sum_{u\in\{n,e\}} \mathbf{h}^H_{b,h}\mathbf{W}_b^{\text{A}}\mathbf{w}_{b,u}^{\text{D}}x_{b,u}[t] + \mathbf{n}_{\text{b,h}}[t],
\end{equation}
where $x_{b,u}[t]$ represents the data symbol transmitted from BS $b$ to UE $u$ and $\mathbf{n}_{\text{b,h}}[t]$ represents the noise at time instance $t$. The receiver at the FD-HAPS is equipped with a hybrid combiner such that the digital combiner weights are $\mathbf{C}^D = [\mathbf{c}^{\text{D}}_{1}, \mathbf{c}^{\text{D}}_{2}] \in \mathbb{C}^{2 \times 2}$ and analog combiner weights are $\mathbf{C}^A = [\mathbf{c}^{\text{A}}_{1}, \mathbf{c}^{\text{A}}_{2}] \in \mathbb{C}^{2 \times M_h}$. The signal received at the HAPS can be represented as
\begin{equation}
\mathbf{y}_{\text{b,h}}[t] = (\mathbf{c}^{\text{D}}_{b})^{H} (\mathbf{C}^{\text{A}})^{H} \mathbf{x}_{\text{b,h}}[t] \label{eq:rec_haps}.
\end{equation}
The transmitter at the FD-HAPS is also equipped with a hybrid beamforming with digital beamforming weights $\mathbf{W}^{\text{D}}_h = [\mathbf{w}^{\text{D}}_{h,1}, \mathbf{w}^{\text{D}}_{h,2}] \in \mathbb{C}^{2 \times 2}$ and analog beamforming weights $\mathbf{W}^{\text{A}}_h = [\mathbf{w}^{\text{A}}_{h,1}, \mathbf{w}^{\text{A}}_{h,2}] \in \mathbb{C}^{M_h \times 2}$. Accordingly, the signal received at the near UE can be derived as combination two data stream from HAPS
\begin{align}
\mathbf{y}_{\text{b,h,$u_e$}}[t] = \mathbf{h}^H_{h,b_n}\sum_{m=1}^{B}\mathbf{W}^{\text{A}}\mathbf{w}_{h,m}^{\text{D}}d_{m,u}[t-\tau]
+ \mathbf{n}_{h,b_n}[t] \label{eq:rec_edge},
\end{align}
where $\tau$ denotes the propagation time delay of the HAPS-UE link caused by the long transmission distance due to the high-altitude deployment of the HAPS. Similarly, the received signal at the near UE can be derived as combination of two datastream from BS and two data stream from HAPS
\begin{align}
\mathbf{y}_{\text{b,h,$u_n$}}[t] &= \mathbf{h}^H_{b,b_e}\sum_{k\in\{n,e\}}\mathbf{W}_b^{\text{A}}\mathbf{w}_{b,k}^{\text{D}}d_{b,u}[t]\notag\\ 
&+ \mathbf{h}^H_{h,b_e}\sum_{m=1}^{B}\mathbf{W}^{\text{A}}\mathbf{w}_{h,m}^{\text{D}}d_{m,e}[t-\tau]
+ \mathbf{n}_{b,b_e}[t] \label{eq:rec_near}.
\end{align}

At the FD-HAPS receiver, signals transmitted by multiple BS overlap on the same frequency, as shown in \eqref{eq:rec_haps}, resulting in inter-BS interference. To separate these superimposed signals, the HAPS applies SIC by sequentially detecting and subtracting the strongest signal components. Therefore, the resulting signal-to-interference-plus-noise ratio (SINR) at the FD-HAPS for BS $b$ is given by

\begin{align}
\gamma^{b,h}_{h} =
\frac{%
|(\mathbf{c}^{\text{D}}_{b})^{H} (\mathbf{C}^{\text{A}})^{H} 
\mathbf{h}^H_{b,h} \mathbf{W}^{\text{A}}_{b} \mathbf{w}^{\text{D}}_{b,e}|^2%
}{%
\begin{aligned}[t]
&|(\mathbf{c}^{\text{D}}_{b})^{H} (\mathbf{C}^{\text{A}})^{H} 
\mathbf{h}^H_{b,h} \mathbf{W}^{\text{A}}_{b} \mathbf{w}^{\text{D}}_{b,n}|^2 + \\
& \sum_{m> b}^B \sum_{u\in\{n,e\}} |(\mathbf{c}^{\text{D}}_{m})^{H} (\mathbf{C}^{\text{A}})^{H} 
\mathbf{h}^H_{m,h} \mathbf{W}^{\text{A}}_{m} \mathbf{w}^{\text{D}}_{m,u}|^2 + \sigma_{b,h}^2
\end{aligned}%
}.
\label{eq:sinr_bs_b_2_haps}
\end{align}

Considering the received signal at the edge UE in (\ref{eq:rec_edge}), the data streams transmitted by the HAPS over the same frequency interfere with each other. To mitigate this interference, SIC is employed at the edge UE by first detecting and canceling the strongest signal component in the received signal. Therefore the corresponding SINR can be expressed as

\begin{eqnarray}
    \gamma_{b}^{h,b_e}=\frac{|\mathbf{h}^H_{h,b_{e}}\mathbf{W}^{\text{A}}\textbf{w}^{\text{D}}_{h,b}|^2}{\sum_{m> b}^B|\mathbf{h}^H_{h,m_{e}}\mathbf{W}^{\text{A}}\textbf{w}^{\text{D}}_{h,m}|^2 + \sigma_{h,b_e}^2}
    \label{eq:SINR_bs_b_2_bs}.
\end{eqnarray}

From (\ref{eq:rec_near}), the near UE receives signals from both the BS and the HAPS, where the HAPS transmission and the edge UE signal from the BS act as interference. The edge UE signal from the BS is first decoded and canceled using SIC. For successful cancellation, the near UE must decode the edge UE signal of its serving BS while treating its own signal as interference. Therefore, the SINR of the edge UE signal at the near UE is given by
\begin{equation}
\gamma_b^{n,e} =
\frac{
\bigl| \mathbf{h}_{b,b_n}^H \mathbf{W}_b^{\text{A}} \mathbf{w}_{b,e}^{\text{D}} \bigr|^2
}{
\bigl| \mathbf{h}_{b,b_n}^H \mathbf{W}_b^{\text{A}} \mathbf{w}_{b,n}^{\text{D}} \bigr|^2
+ 
\displaystyle\sum_{\substack{m=1 \\ m \neq b}}^B 
\bigl| \mathbf{h}_{h,m_n}^H \mathbf{W}^{\text{A}} \mathbf{w}_{h,m}^{\text{D}} \bigr|^2
+ \sigma^2_{n,e}
}.
\end{equation}
Once the edge UE signal is subtracted from received signal, the remaining interference signal is due to edge UE signal from the HAPS corresponding to the other BS. Therefore, the SINR of the received signal at near UE can be derived as
\begin{flalign}
    \gamma_{b}^{b,b_n} = \frac{|\mathbf{h}^H_{b,b_{n}}\mathbf{W}_b^{\text{A}}\mathbf{w}_{b,n}^{\text{A}}|^2}{\sum_{m=1, m\neq b}^B|\mathbf{h}^H_{h,b_n}\mathbf{W}^{\text{A}}\textbf{w}^{\text{A}}_{h,m}|^2 +\sigma^2_{b,b_n}}.
\end{flalign}

From the above SINR equations, the overall SINR of edge UE can be obtained as
\begin{align}
    \gamma_{b}^{b,b_e} = \min(\gamma^{h,b_e}_{b}, \gamma^{b,h}_{h}, \gamma_{b}^{n,e}).
\end{align} 
The achievable rates for the near UE $n$ and edge UE $u$ of BS $b$ can be obtained as,
\begin{eqnarray}
R_{b,u} = \log_{2}(1 + \gamma_{b}^{b,b_n}) \text{  for u} \in \{n,e\}.
\end{eqnarray} 

\section{Problem Formulation} \label{sec:problem_formulation}

In this work, we are interested in obtaining a problem to jointly optimize the transmit hybrid precoders at the BS and FD-HAPS, the receiver combiner at the FD-HAPS such that the EE ($\eta$) of the system is maximized with the minimum rate ($r_{\text{QoS}}$) constraints and the maximum transmit power constraints at the BS and the FD-HAPS. \textcolor{black}{EE is defined as the ratio of the sum rate achieved by all users and the sum transmit power consumed by the base stations and the HAPS. It represents the number of information bits successfully delivered per unit of transmit energy.} To formulate the EE maximization problem, we start with defining the EE ($\eta$) as follows:
\begin{eqnarray}
    \eta = \frac{\sum_{b=1}^B \sum_{u \in\{n,e\}} R_{b,u}}{\sum_{b=1}^B\sum_{u \in \{n,e\}}|\mathbf{W}_b^{\text{A}}\mathbf{w}_{b,u}^{\text{D}}|_F^2 + \sum_{b=1}^B|\mathbf{W}^{\text{A}}_h\mathbf{w}_{h,b}^{\text{D}}|_F^2}.
\end{eqnarray} 

Therefore, the beamforming optimization problem can defined as
\begin{IEEEeqnarray*}{lcl}\label{eq:Orig_P1}
&\underset{\substack{\mathcal{W}^{\text{A}}, \mathcal{W}^{\text{D}}, \mathbf{W}^{\text{A}}_h,\\
\mathbf{W}^{\text{D}}_h, \mathbf{C}^{\text{A}}, \mathbf{C}^{\text{D}}}}{\max}\,\, & \eta  \IEEEyesnumber \IEEEyessubnumber* \label{eq:P1_Obj}\\
&\text{s.t.} & R_{b,u}\geq R_{\text{min}}, \forall b, u, \label{eq:P1_far_UE_rate_constr}\\
&& \left|\mathbf{W}_b^{\text{A}}\right|_{i,j}^2 = \frac{1}{M_b},\, \forall i, j, b, \label{eq:P1_BS_const_mod_constr}\\
&& \left|\mathbf{W}^{\text{A}}\right|_{i,j}^2 = \frac{1}{M_{\text{h}}}, \, \forall i, j, \label{eq:P1_HIBS_const_mod_constr}\\
&& \left|\mathbf{C}^{\text{A}}\right|_{i,j}^2 = \frac{1}{M_{\text{h}}},\, \forall i,j,\label{eq:P1_HIBS_combiner_const_mod_constr}\\
&& \sum_{u \in\{n,e\}}
\left\|\mathbf{W}_b^{\text{A}}\mathbf{w}_{b_u}^{\text{D}}\right\|_F^2
\leq P_{\max}^b,\quad \forall b, \label{eq:P1_BS_mat_product_constr}\\
&& \sum_{b=1}^B
\left\|\mathbf{W}^{\text{A}}_h\mathbf{w}_{h,b}^{\text{D}}\right\|_F^2
\leq P_{\max}^h, \label{eq:P1_HIBS_mat_product_constr}
\end{IEEEeqnarray*}
 where $\mathcal{W}^{\text{A}} = [\mathbf{W}^A_1,...,\mathbf{W}^A_B]$, $\mathcal{W}^{\text{D}} = [\mathbf{W}^D_1,...,\mathbf{W}^D_B]$, $\mathbf{w}^{\text{A}}_{h,b}$ and $\mathbf{w}^{\text{D}}_{h,b}$ are the analog and digital precoders at the FD-HAPS, $\mathbf{C}^{\text{A}}$, $\mathbf{C}^{\text{D}}$ are the analog and digital combiners at the FD-HAPS, respectively. The constraint \eqref{eq:P1_far_UE_rate_constr} is used to ensure that all UE achieve a minimum rate of $R_{min}$. The constraints \eqref{eq:P1_BS_const_mod_constr}-\eqref{eq:P1_HIBS_combiner_const_mod_constr} is used to ensure that analog beamforming weights and combiner weights have constant modulus and it is equal to inverse of number of antenna elements. These constraints reflect the practical realization of analog beamforming which uses phase shifters and only changes the phase of the signal and not its strength. Setting a constant value keeps the total transmitted power under control.
 Constraints \eqref{eq:P1_BS_mat_product_constr} and \eqref{eq:P1_HIBS_mat_product_constr} ensures that transmit power of BS and HAPS are within the maximum power limits of $P_{max}^b$ and, $P_{max}^h$
respectively.

\section{Proposed solution} \label{sec:proposed_solution}
\subsection{Analog beamforming weights design} \label{sec:analog_solution}
Joint optimization of the hybrid beamforming and combiner at the BS and the FD-HAPS is challenging, due to non-convex objective function, matrix-product power constraints \eqref{eq:P1_BS_mat_product_constr} and \eqref{eq:P1_HIBS_mat_product_constr}, and constant-modulus constraints \eqref{eq:P1_BS_const_mod_constr}, \eqref{eq:P1_HIBS_combiner_const_mod_constr} and \eqref{eq:P1_HIBS_const_mod_constr}. To this end, we first design the analog beamforming weights using a low-complexity strategy that directly maps the phase of the complex-conjugate channel coefficients onto the beamforming weights, thereby maximizing array gain while satisfying the constant-modulus constraints. Accordingly, the analog beamforming weights associated with the near UE of BS $b$ is expressed as
\begin{align}
\mathbf{w}^{\text{A}}_{b,n} =& [e^{-j\angle [\mathbf{h}^{*}_{b,b_n}]_1}, e^{-j\angle [\mathbf{h}^{*}_{b,b_n}]_2}, \ldots, e^{-j\angle [\mathbf{h}^{*}_{b,b_n}]_{M_b}}]^T, \forall b \label{eq:RF_precoder_near_UE},
\end{align}
where $[\mathbf{h}_{b,b_n}]_i$ is the $i$th element of the channel vector $\mathbf{h}_{b,b_n}$. For the edge UE served by BS $b$, the analog beamforming weights is constructed based on the transmit array response vector corresponding to the dominant departure direction, expressed as
\begin{eqnarray}
\mathbf{w}^{\text{A}}_{b,e} = \mathbf{r}_t(\theta^{D,0}_{b,\text{h}}, \varphi^{D,0}_{b,\text{h}}),\quad \forall b. \label{eq:RF_precoder_near_UE}
\end{eqnarray}

Similarly, we can obtain the analog beamforming weights of FD-HAPS based on the channel from HAPS to edge UE as
{\begin{align}
\mathbf{w}^{\text{A}}_{h,b} =& [e^{-j\angle [\mathbf{h}^{*}_{h,b_e}]_1}, e^{-j\angle [\mathbf{h}^{*}_{h,b_e}]_2}, \ldots, e^{-j\angle [\mathbf{h}_{h,b_e}]_{M_{\text{h}}}}]^T, \forall b.\label{eq:RF_precoder_far_UE}
\end{align}
Further, we can obtain the analog combiner weights of FD-HAPS using the receive array response vector as
\begin{eqnarray}
\mathbf{c}^{\text{A}}_b = \mathbf{r}_r(\theta^{D,0}_{b,\text{h}}, \varphi^{D,0}_{b,\text{h}}),\quad \forall b. \label{eq:RF_precoder_near_UE}    
\end{eqnarray}
Since the analog beamforming and combiner weights are known, \eqref{eq:Orig_P1} can be reduced to the following problem
\begin{IEEEeqnarray}{lcl}\label{eq:P2_digi}
\underset{\substack{\mathcal{W}^{\text{D}}, \mathbf{W}^{\text{D}}_h, \mathbf{C}^{\text{D}}}}{\max} 
& & \eta \IEEEyesnumber \IEEEyessubnumber* \label{eq:P2} \\
\text{s.t.} 
& & \eqref{eq:P1_far_UE_rate_constr},\,
\eqref{eq:P1_BS_mat_product_constr}-\eqref{eq:P1_HIBS_mat_product_constr}.
\end{IEEEeqnarray}

\subsection {Digital beamforming weights design} \label{sec:digital_solution}
Next, we focus on designing the digital beamforming weights by solving \eqref{eq:P2_digi}. This step is still challenging due to the non-convex objective function and rate constraints \eqref{eq:P1_far_UE_rate_constr}. Conventional iterative methods, such as ACE \cite{Hassan-Shabih-Alamari-Sultan-Usama-Electronics-2023} and stochastic beamforming \cite{Viswanath-Tse-IT-2002}, rely on model-based updates and often struggle to scale effectively in dense networks with high interference due to large number of UE \cite{ZhangTansu-TWC-2019},\cite{Lu-Zhao-Shi-Globecomm-2020}. To overcome these limitations, a GNN-based optimization framework is adopted to learn a direct nonlinear mapping from instantaneous channel state information to digital beamforming weights. 


\subsubsection{GNN and Wireless Networks} \label{sec:nn_gnn}
GNN are specialized neural network architectures designed to learn from graph-structured data, where nodes (vertices) are connected by edges (links). Unlike traditional neural networks that handle independent feature vectors, GNNs model relationships among nodes by iteratively exchanging and aggregating information through a message-passing mechanism \cite{Scarselli-Franco-Gori-Marco-TNN-2009,Wu-Pan-Fengwen-Guodong-Chengqi-Philip-TNN-2021}. Each node has a state vector, and each edge has a feature vector, facilitating complex dependency learning among nodes.

Wireless communication networks can leverage GNNs, as they naturally represent their elements as graphs—network elements are nodes and wireless links are edges. GNNs have been effectively used in these networks to optimize resource allocation, power control, and beamforming due to their ability to utilize spatial relationships and interference patterns \cite{JIANG-CC-2022, Lim-Vu-SSP-GNNBF-2023, Ivanov-Antoni-Tonchev-Poulkov-Manolova-ISTC-2022}. In this work, we propose a GNN-based architecture to optimize digital beamforming weights, modeling base stations (BS), high-altitude platform stations (HAPS), and user equipment (UE) as nodes, with wireless channels and interference as links. This architecture includes three neural networks: Node NN ($NN_{\text{node}}$), Link NN ($NN_{\text{link}}$), and Beamforming NN ($NN_{\text{beam}}$). Let $\mathbf{v}_t$ represent the state vector for each node, and $\mathbf{H}_{r,t}$ denote the link feature between nodes $t$ and $r$.

The Link NN generates message vector for the link between nodes $t$ and $r$ as $\mathbf{m}_{r,t} = NN_{\text{node}}(\mathbf{v}_t, \mathbf{H}_{r,t})$. It essentially represents the signal transmitted from node $t$ that has propagated through the channel $\mathbf{H}_{r,t}$ to reach node $r$. Similarly, all other messages are created for other links. In the next step, the receiving nodes send messages to Node NN and update their state vector as $\mathbf{v}_{r} = NN_{\text{link}}(\mathbf{m}_{r,1},\ldots, \mathbf{m}_{r,T}, \mathbf{v}_r)$, where $T$ denotes the number of neighboring nodes sending messages to node $r$.


The updated state vectors associated with the BS and the HAPS nodes are then fed to the Beamforming NN to generate the digital beamforming and combining weights as $\{\mathcal{W}^{\text{D}}, \mathbf{W}^{\text{D}}_h, \mathbf{C}^{\text{D}}\} = NN_{\text{beam}} (\mathbf{v}_{1}, \ldots , \mathbf{v}_{B}, \mathbf{v}_{h})$.




\subsubsection{Problem Transformation to GNN} \label{sec:penalty}

In general, NN training requires an objective function to be optimized. For the beamforming optimization considered here, we instead employ a penalty function derived from~\eqref{eq:P2_digi}. During training, this penalty is minimized while enforcing all constraints. The penalty function is defined as
\begin{flalign}\label{eq:Penalty}
\text{$\rho$} = &\frac{\lambda_1}{EE} + \lambda_2 R_c + \lambda_3 P_c,
\end{flalign}
\begin{flalign} \label{eq:Rc}
&R_c = \sum_{b=1}^B \sum_{u\in\{n,e\}} \max\left(0, R_{\text{min}} - R_{b,u}\right),&\\
    &P_c = \sum_{b=1}^B\max(0, P^b - P_{\text{max}}^b) + \max(0, P^h - P_{\text{max}}^h),& \label{eq:Pc} \\
&{P^b} = \sum_{u\in\{n,e\}} \left|\mathbf{W}_b^{\text{A}}\mathbf{w}_{b,u}^{\text{D}}\right|_F^2, \forall b,& \\
&{P^h} = \sum_{b=1}^B\left|\mathbf{W}^{\text{A}}_h\mathbf{w}_{h,b}^{\text{D}}\right|_F^2,&
\end{flalign}
where, $R_c$ represents the minimum achievable rate constraint described in \eqref{eq:P1_far_UE_rate_constr} and $P_c$ is maximum power constraints described in  \eqref{eq:P1_BS_mat_product_constr} and \eqref{eq:P1_HIBS_mat_product_constr}. The terms $P^h$ and $P^b$ represent the transmit powers of HAPS and BS $b$, respectively. $\lambda_1$, $\lambda_2$, $\lambda_3$ are the scaling factors. We consider $\lambda_2$ and $\lambda_3$ to higher than $\lambda_1$ to ensure that the minimum rate and maximum power constraints are strictly enforced. This prevents the GNN from prioritizing EE maximization at the expense of violating the minimum rate and maximum power constraints. Based on the penalty function, the final problem, equivalent to \eqref{eq:P2}, can be defined as
\begin{align}
    &\underset{\substack{\mathcal{W}^{\text{D}}, \mathbf{W}^{\text{D}}_h, \mathbf{C}^{\text{D}}}}{\min}\,\, \text{$\rho$} \label{eq:P3}.
\end{align}
The optimization problem in \eqref{eq:P3} is equivalent to \eqref{eq:P2} and is solved using the penalty formulation given in \eqref{eq:Penalty}. The first term in the penalty function is defined as the inverse of the EE. When the penalty is minimized, the first term is reduced, which increases the EE and achieves the main objective of \eqref{eq:P2}. The term $R_c$ enforces the minimum rate constraint in \eqref{eq:P1_far_UE_rate_constr}. This term is zero when the achievable rate $R_{b,u}$ satisfies the QoS requirement $R_{\text{min}}$, and becomes positive when the rate falls below this threshold, increasing the penalty. The term $P_c$ accounts for the transmit power constraints in \eqref{eq:P1_BS_mat_product_constr} and \eqref{eq:P1_HIBS_mat_product_constr}. It remains zero as long as the transmit powers are within the allowable limits $P^b_{\text{max}}$ and, $P^h_{\text{max}}$ and increases when these values are above thresholds. As the optimization minimizes the penalty, it naturally drives the second and third terms toward zero, ensuring that the rate constraint and power constraint are satisfied.

\begin{figure}[h] 
    \centering    \includegraphics[width=1.0\linewidth]{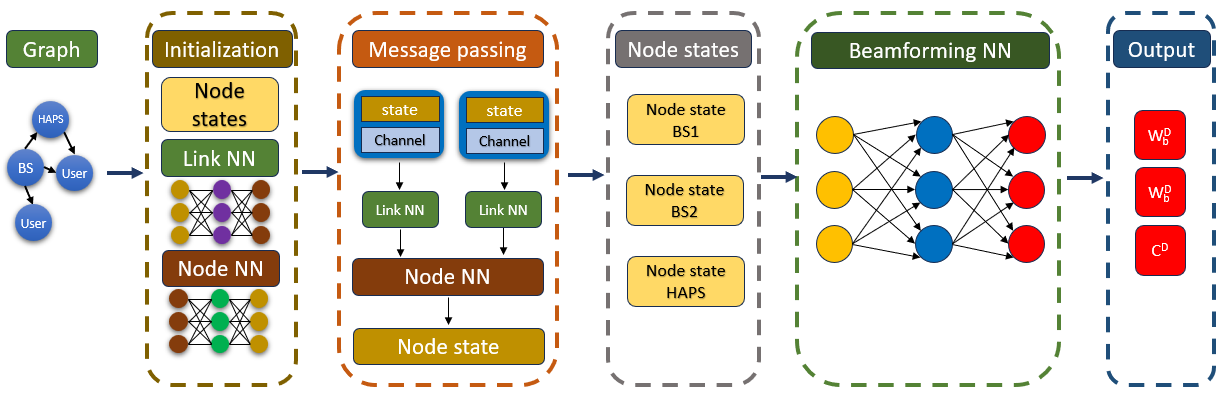} 
    \caption{Proposed GNN-based optimization architecture}
    \label{fig:gnn}
\end{figure}

\begin{figure}[h]  
    \centering
    \includegraphics[width=1.00\linewidth]{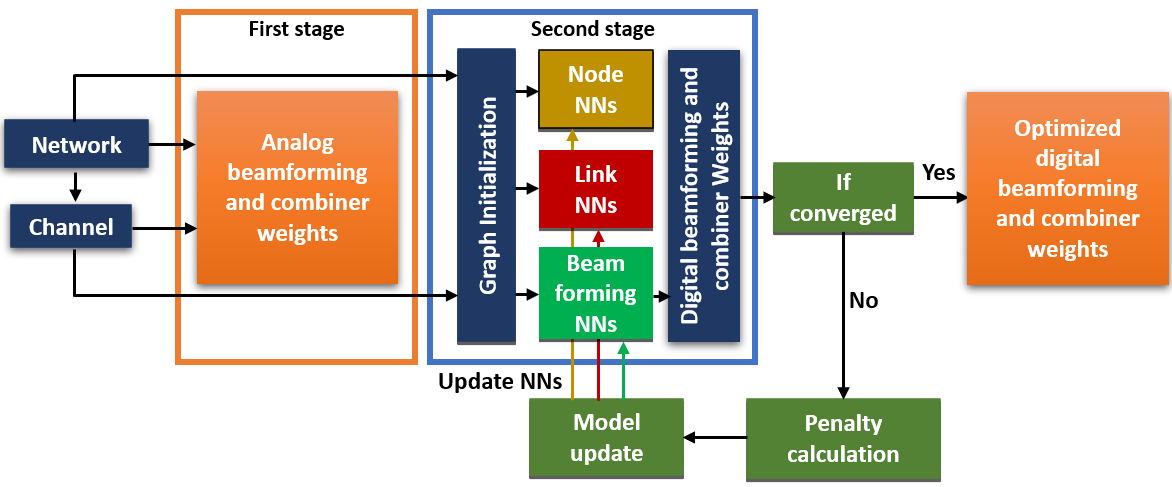} 
    \caption{Flow diagram of GNN optimization}
    \label{fig:online_flow}
\end{figure}
\textcolor{black}{Fig. \ref{fig:gnn} illustrates the GNN architecture that models the network as a graph with BS, HAPS, and UE as nodes, and wireless links. Fig. \ref{fig:online_flow} is a detailed representation of second stage in Fig. \ref{fig:gnn}.} The GNN is executed centrally at the HAPS, which collects the required channel information from the BS and UE and performs the message-passing computations. As shown in the \textcolor{black}{Fig. \ref{fig:gnn}}, the GNN employs three components: Link NNs use channel features and learn propagation effects, Node NNs aggregate the output of Link NN to capture signal and interference patterns at receivers, and a Beamforming NN generates digital beamforming and combining weights from the node states. \textcolor{black}{Fig. \ref{fig:online_flow}} represents the complete workflow utilizing this architecture. Starting with network configuration and channel matrix, the first stage computes analog weights using closed-form solutions based on channel phases for near users and array steering for HAPS-assisted edge users. The second stage constructs the graph from \textcolor{black}{Fig. \ref{fig:online_flow}} and iteratively optimizes digital weights: the GNN forward pass generates beamforming weights, and a penalty function is computed to measure EE and constraint satisfaction. The gradients are calculated and the NN are updated using backpropagation technique. This loop continues until maximum iterations, yielding optimized hybrid beamforming that maximizes energy efficiency while satisfying QoS and power constraints.
\begin{table}[t]
\centering
\caption{Simulation Parameters}
\begin{tabular}{|l|l|}
\hline
\textbf{Parameter} & \textbf{Value} \\ \hline
No. of antennas in BS & $M_{\mathrm{bh}}=8,\; M_{\mathrm{bv}}=8$ \\ \hline
No. of antennas in HAPS & $M_{\mathrm{hx}}=10,\; M_{\mathrm{hy}}=10$ \\ \hline
Carrier frequency & $f_c = 28$ GHz \\ \hline
Bandwidth & $B = 500$ kHz \\ \hline
Cell radius & $200\text{ m},\; 1000\text{ m}$ \\ \hline
HAPS height & $22$ km \\ \hline
Maximum transmit power & $P_{\mathrm{hx}} = 45$ dBm,\; $P_b = 45$ dBm \\ \hline
No. of NLoS paths & $S = 5$ \\ \hline
No. of hidden layers & DNN $=2$, CNN $=2$, GNN $=2$ \\ \hline
Neurons per hidden layer & DNN: $64, 32$;\; CNN: $64, 32$;\; GNN: $32, 32$ \\ \hline
Scaling factors & $\lambda_1 = 1,\; \lambda_2 = 10,\; \lambda_3 = 10$ \\ \hline
\end{tabular}\label{table:sim_params}
\end{table}

\section{Simulation Results and Discussion} \label{sec:sim}
\subsection{Simulation Setup}

In this section, we evaluate the performance of the proposed GNN-based online optimization framework. To enable a fair comparison among optimization-driven baselines, topology-agnostic learning, and structure-aware learning, we consider three benchmark methods: (i) an ACE-based optimizer that iteratively updates a sampling distribution over candidate beamformers to maximize EE \cite{Hassan-Shabih-Alamari-Sultan-Usama-Electronics-2023}; (ii) a fully connected DNN that learns a direct mapping from CSI (and network descriptors) to beamforming parameters \cite{Hu-Rentao-IEEEAccess-2021}; and (iii) a CNN-based model that applies convolutional and pooling layers prior to the fully connected layers to exploit local feature structure \cite{Ayad-Mohammed-Fethi-IJECE-2025}.

For each network realization, the instantaneous CSI and network topology are used to construct the graph input and initialize the NN architecture described in Section \ref{sec:nn_gnn}. The GNN parameters are then updated online by minimizing the penalty-based objective in Section \ref{sec:penalty} via backpropagation, following the workflow in \textcolor{black}{Fig. \ref{fig:online_flow}}. To study scalability and coverage effects, \textcolor{black}{we consider two cell radii, $200$~m and $1000$~m.} Also, for each realization, the optimizer is executed for at most 1000 iterations with early stopping enabled. The list of simulation parameters is summarized in Table \ref{table:sim_params}. Furthermore, to ensure a fair comparison, all schemes are evaluated under identical channel realizations and system parameters.

\subsection{Results and Discussion} \label{sec:results}
\begin{figure}[!h]
    \centering
    \begin{subfigure}[b]{0.492\columnwidth} \includegraphics[width=\linewidth]{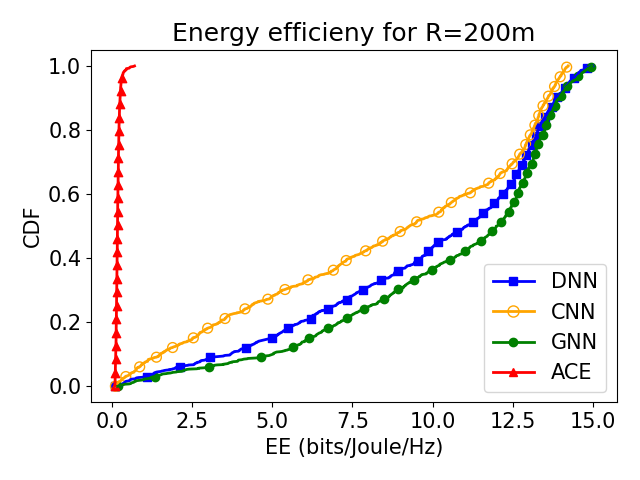}  
    \end{subfigure}
    \begin{subfigure}[b]{0.492\columnwidth}
    \includegraphics[width=\linewidth]{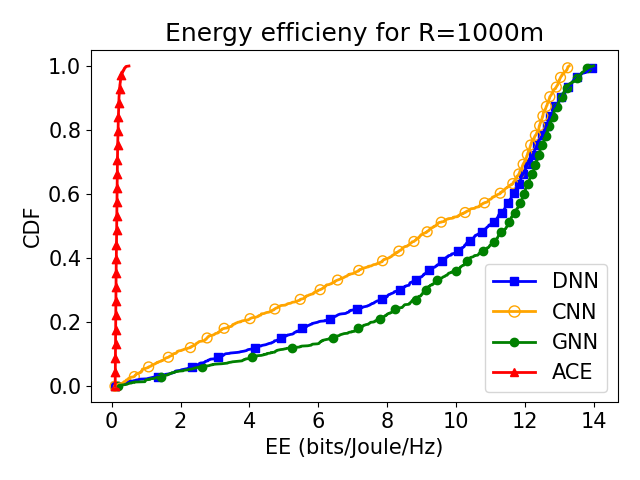}
    \end{subfigure}
    \caption{CDF vs. EE for different solution approaches and cell radii $\in \{200, 1000\}$ m.}
    \label{fig:online_ee}
\end{figure}

We compare the EE performance using the CDF for different cell radii, such as 200 m and 1000 m, to examine the impact of increasing cell size on EE. The results show that, based on the observed CDF behavior, the EE performance does not decrease significantly as the cell radius increases.
Fig. \ref{fig:online_ee} plots the cumulative distribution functions (CDFs) versus EE, obtained using the GNN-based approach, under different terrestrial cell radii and compares its performance with three benchmarks approaches (i.e., ACE, DNN, CNN).  It can be observed that the GNN-based approach consistently achieves higher EE across all considered cell radii. The higher EE is due to the ability of the GNN to explicitly capture inter-cell interference and coordination through graph-based message passing. Although DNN and CNN outperform the ACE approach, their lack of explicit topology awareness limits their performance. The ACE method yields the lowest energy efficiency, while the GNN provides the best overall performance. 

\begin{figure}[!h]
    \centering
    \begin{subfigure}[b]{0.492\columnwidth}
        \includegraphics[width=\linewidth]{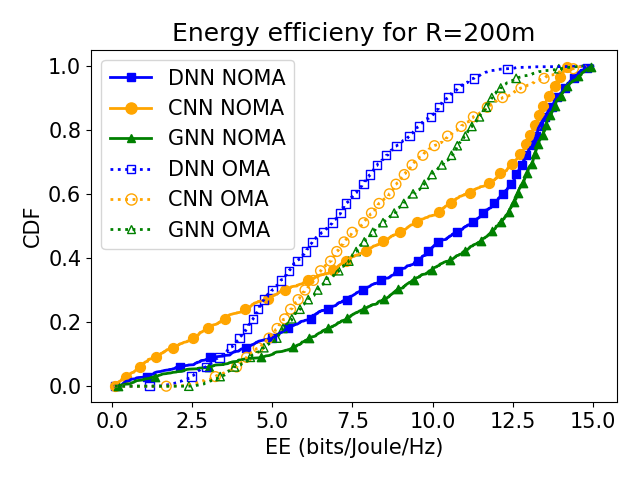}  
    \end{subfigure}
    \begin{subfigure}[b]{0.492\columnwidth}
        \includegraphics[width=\linewidth]{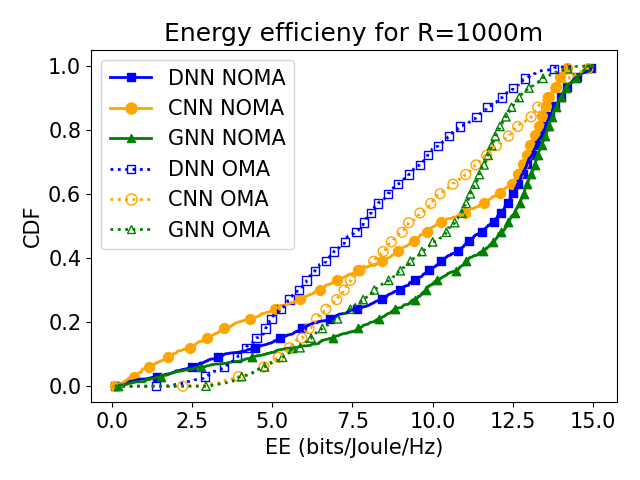}
    \end{subfigure}
    \caption{CDF vs. EE with NOMA and OMA techniques across different cell radii $\in \{200, 1000\}$ m.}
    \label{fig:online_ee_oma}
\end{figure}
Fig. \ref{fig:online_ee_oma} compares the CDF versus EE performance of NOMA and orthogonal multiple access (OMA) techniques. The results show that NOMA consistently achieves higher EE than OMA for all cell radii, highlighting its superior energy efficiency in multi-user scenarios. As the cell radius increases, the overall EE slightly decreases for all schemes due to higher path loss and reduced spectral efficiency at larger distances. Among the techniques, the proposed GNN-based methods outperform CNN- and DNN-based approaches for both NOMA and OMA, demonstrating the GNN’s ability to effectively capture interference patterns and optimize resource allocation. The CDF curves further show that the GNN-based NOMA achieves higher EE for a larger fraction of users, with the curves shifting right and steeper compared to other methods, indicating more consistent and robust EE gains. When comparing performance across two different cell radii, a significant variation is observed in the GNN performance under OMA, whereas the deviation is considerably smaller in the NOMA case. This demonstrates the consistency of the proposed GNN-based technique when integrated with the NOMA framework.

\begin{figure}[h!]
    \centering
    \begin{subfigure}[b]{0.492\columnwidth}
        \includegraphics[width=\linewidth]{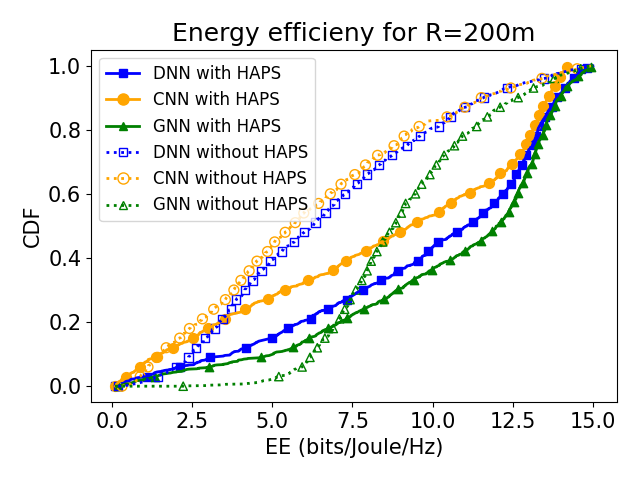}  
    \end{subfigure}
    \begin{subfigure}[b]{0.492\columnwidth}
        \includegraphics[width=\linewidth]{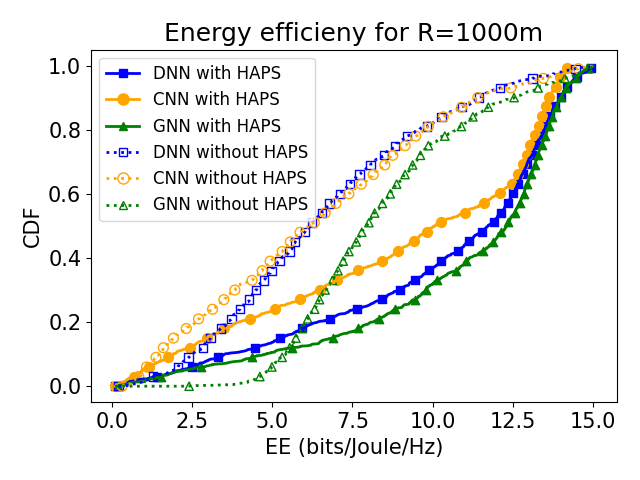}
    \end{subfigure}
    \caption{CDF graphs of EE with and without HAPS across different cell radii $\in \{200, 1000\}$ m.}
    \label{fig:online_ee_no_haps}
\end{figure}
Fig. \ref{fig:online_ee_no_haps} compares the CDF versus EE performance and studies the effect of using HAPS in the network. Higher EE is achieved when HAPS is used in all three solution approaches to that of without HAPS. This observation highlights that the HAPS-assisted NTN help in improving the EE of the network. Further, the proposed GNN-based method achieves a 5th-percentile EE higher than the other three solution approaches. This significant improvement demonstrates that the GNN maintains strong EE performance even under challenging conditions, such as severe interference and low SINR, while ensuring improved capacity for edge users through the HAPS relay, highlighting the robustness of the proposed approach. When comparing performance across two different cell radii, a significant variation is observed in the GNN performance without HAPS, whereas the deviation is considerably smaller in the HAPS-assisted case. This demonstrates the consistency of the proposed GNN-based technique when integrated with HAPS.

\section {Conclusion} \label{sec:conclusion}
In this work, we proposed an online beamforming optimization approach using GNN that enhances the energy-efficient transmission in HAPS-assisted multi-cell NOMA networks. The performance of the proposed GNN-based solution is evaluated and compared with DNN, CNN, and ACE-based approaches. Numerical results show that NN can effectively address the non-convex and coupled beamforming problem without relying on offline training. Among all methods, the GNN-based approach consistently achieved higher EE and more stable due to its topology-aware graph representation and message-passing mechanism. However, DNN and CNN provided moderate improvements but were less effective in exploiting network structure, while the ACE method yielded lower EE. Furthermore, the network with HAPS shown to provide higher EE and a 5th percentile as compared to one without HAPS. These findings highlight the importance of HAPS and topology-aware learning for online optimization in multi-cell NOMA systems. Future work will consider extensions to multiple HAPS deployments, user mobility, and imperfect channel state information.

\bibliography{ref.bib}
\bibliographystyle{IEEEtran}

\end{document}